\documentclass[aps,amsfonts,preprintnumbers,amsmath,amssymb,nofootinbib,tightenlines,preprint,showpacs]{revtex4}

\usepackage{bm}
\usepackage{epsfig}
\usepackage{subfigure}
\usepackage[latin1]{inputenc}
\usepackage{float}
\usepackage{graphicx}
\usepackage{cancel}
\usepackage{mathrsfs}
\usepackage{amssymb}
\usepackage{amsfonts}
\usepackage{amsmath}
\usepackage{slashed}

\begin{document}

\title{Constraints on the excitations \\ in the Strongly Coupled Standard Model}

\author{Francesco D'Eramo}
\email{fderamo@mit.edu}
\affiliation{Center for Theoretical Physics,\\ Massachusetts Institute of Technology, \\ Cambridge, MA 02139}

\preprint{MIT-CTP 4034}

\begin{abstract}
The Strongly Coupled Standard Model predicts a rich spectrum of excited states at the Fermi scale. We study the first radial excitations of the vector bosons. The inclusion of these new states affects the low energy phenomenology of the model. We put constraints on the effective couplings by performing a global fit with the electroweak observables, and we find that the excitations have to be rather decoupled from the low-energy states.
\end{abstract}
\maketitle

\section{Introduction}
\label{intro}
The Large Hadron Collider (LHC) will start soon its operations, making accessible for the first time the energy range above the Fermi scale $G_{F}^{-1/2}$. Electroweak symmetry is broken at this scale, and the mechanism which causes ElectroWeak Symmetry Breaking (EWSB) is still unknown. In the Standard Model (SM) the electroweak symmetry is broken by the vacuum expectation value (vev) of a scalar field, whose associated particle, the Higgs boson, has not been observed yet. However the agreement with the experimental data is excellent. Despite this tremendous success, the SM is not believed to be a fundamental theory of nature, since the Higgs potential in unstable under radiative corrections and very sensitive to higher energy scales, leading to the well known hierarchy problem. This suggests that we should expect new physics at the Fermi scale replacing the SM, and many candidate models have been proposed in the last thirty years, as supersymmetry \cite{Veltman:1976rt,Witten:1981nf,Dimopoulos:1981zb,Martin:1997ns}, technicolor \cite{Weinberg:1975gm,Weinberg:1979bn,Susskind:1978ms}, Little Higgs \cite{ArkaniHamed:2001nc,Schmaltz:2005ky}, large extra dimensions \cite{ArkaniHamed:1998rs,Antoniadis:1998ig,ArkaniHamed:1998nn,Cremades:2002dh,Kokorelis:2002qi}, warped extra dimensions \cite{Randall:1999ee,Randall:1999vf}, holographic models \cite{Csaki:2003zu,Contino:2003ve,Nomura:2003du,Agashe:2004rs}. A nice overview can be found in references \cite{Giudice:2007qj,Barbieri:2008xg}.

In this work we reconsider an alternative and very specific model for the electroweak interactions, which predicts a rich spectrum of particles at the Fermi scale, namely the Strongly Coupled Standard Model (SCSM) \cite{Abbott:1981re,Abbott:1981yg,Claudson:1986ch}. The SCSM is the confining version of the usual $SU(2)_{L} \times U(1)$ weak-coupling Standard Model. In the usual SM the mass scale of the weak interaction is given by the vev of a scalar field, which also cuts off the IR growth of the $SU(2)_{L}$ gauge coupling. In the SCSM the $SU(2)_{L}$ gauge theory is assumed confining rather than spontaneously broken: the gauge coupling becomes large at the scale $\Lambda_2 \simeq G_{F}^{-1/2}$ and the vev of the scalar field is vanishingly small. The SCSM Lagrangian is the same as the conventional $SU(2)_{L} \times U(1)$ SM Lagrangian, with the usual particle content and quantum number assignments. The only changes are the values of the $SU(2)_{L}$ coupling constant and the vacuum expectation value of the scalar field. Since the $SU(2)_{L}$ gauge interaction is confining the physical particles are composite bound into $SU(2)_{L}$ singlets, and the observed `weak interactions' are the residual interactions between the composite particles, as the strong interactions between color singlets in QCD. It has already been shown that this confining theory:
\begin{itemize}
\item has a spectrum which matches the SM spectrum \cite{Banks:1979fi,Fradkin:1978dv,Dimopoulos:1980hn};
\item has a low-energy charged and neutral current weak interactions experimentally indistinguishable from those of the SM if we assume vector meson dominance (VMD) \cite{Abbott:1981re,Abbott:1981yg};
\item reproduces the correct $W$ and $Z$ phenomenology (under certain dynamical assumptions) \cite{Claudson:1986ch}.
\end{itemize}
A few comments are in order. The composite fermions have form factors characterized by the scale $\Lambda_2$, and since they are bound by an asymptotically free interaction the form factors have to vanish for infinite momentum transfer. The VMD hypothesis consists in the assumption that the electromagnetic vector form factor for the fermions is saturated only by the $W$ boson, as we will show in the next section. In order to reproduce the observed $W$ and $Z$ phenomenology we have also to make three dynamical assumptions: chiral-symmetry breaking is prevented by the scalar field; the effective fermion-antifermion-$W$ coupling is small; the $W$ and $Z$ masses are smaller than $\Lambda_2$. Given these assumptions the low-energy effective Lagrangian of the SCSM is indistinguishable from the SM Lagrangian. However, at energy well above the $W$ mass deviations from the SM predictions are expected and a rich spectrum of new particles should be observed. These resonances affect the low-energy physics through their contributions to the electroweak observables, therefore while we wait for the LHC to directly look for these particles we can constrain the model by using the available experimental data.

We study the first radial excitations of the $W^{i}$ vector bosons, the $W^{\prime\,i}$ triplet, by including these resonances into the effective theory and by constraining the effective couplings. The inclusion of this triplet violates the VMD hypothesis, which is one assumption we need to reproduce the Standard Model. Given the excellent agreement between the experimental data and the Standard Model we expect this hypothesis to be correct up to a very good accuracy, and then these resonances to be highly constrained by the ElectroWeak Precision Tests (EWPT). Hence the motivation to study these particular excited states is clear: they are the most constrained by the EWPT. Before the LEP measurements at the $Z$ pole the inclusion of this triplet was perfectly consistent with the experiments \cite{Korpa:1985if}. However, the comparison with the first data at the $Z$ pole showed that the effective couplings have to be rather small \cite{Sather:1995tw}. In this paper we update and improve the analysis of \cite{Sather:1995tw}, by making a global fit with all the observables at the $Z$ pole and the $W$ boson mass, and extending the analysis to nonoblique corrections.

The structure of the paper is as follows. In Section \ref{themodel} we review the SCSM, focusing only on the particles of the spectrum we need in order to reproduce the SM (the so called `minimal sector'). In Section \ref{triplet} we include the excited $W^{\prime\,i}$ triplet in the effective theory, and we show that to describe such excitations we need three effective couplings. In Section \ref{efflag} we write the effective Lagrangian we need to get the contributions to the electroweak observables from the excited states, which are computed in Section \ref{EWobs}. The results of the global fit are presented in Section \ref{ResultsFit}. In Section \ref{Higgs} we make again the analysis for a heavy Higgs boson. Conclusions are given in Section \ref{concl}.

\section{The model and its symmetries}
\label{themodel}
The model is based on the usual $SU(2)_{L} \times U(1)$ electroweak Lagrangian with the same matter content. We have $N$ left-handed fermion doublets $\psi^{a}_{L}$, where the index $a$ goes from $1$ to $N$. If we neglect the strong interactions the index $a$ labels colors as well as flavors, then for $n_{f}$ families of quarks and leptons we have $N=4 n_{f}$. Each quark doublet has $U(1)$ charge $y_{a}=1/6$, whereas each lepton doublet has $y_{a}=-1/2$. In addition for each quark doublet there are two right-handed $SU(2)_{L}$ singlets $u_{R}$ and $d_{R}$, with $U(1)$ charge $2/3$ and $-1/3$ respectively, and for each lepton doublet there is one right-handed $SU(2)_{L}$ singlet $e_{R}$ with $U(1)$ charge $-1$\footnote{We are free to include into the spectrum the right-handed neutrinos, which are singlets under the gauge interactions. This does not affect the remaining of our discussion.}. We denote these right-handed fermion fields by $\psi^{b}_{R}$, where the index $b$ goes from $1$ to $7 n_{f}$. There is also a scalar $SU(2)_{L}$ doublet $\phi$, with $U(1)$ charge $-1/2$. We have now all the informations we need to write the Lagrangian of the model, which results in \footnote{We are considering the massless fermions limit. We will discuss how to generate the masses for the physical fermions later.}
\begin{equation}
\mathcal{L}= i\bar{\psi_{L}^{a}}\slashed{D} \psi_{L}^{a} + i\bar{\psi_{R}^{b}}\slashed{D} \psi_{R}^{b} + \frac{1}{2} Tr\left[\left(D_{\mu}\Omega\right)^{\dag}\left(D_{\mu}\Omega\right)\right]-\frac{1}{4} \mathcal{F}_{\mu\nu}\mathcal{F}^{\mu\nu}-\frac{1}{2}Tr\left[G_{\mu\nu}G^{\mu\nu}\right]-\frac{\lambda}{2}\left(Tr\left[\Omega^{\dag}\Omega\right]-2v^2\right)^2
\label{SCSMLag}
\end{equation}
where the covariant derivatives are given by
\begin{equation}
\begin{split}
D_{\mu}\Omega =& \,\partial_{\mu}\Omega - i g_2 \omega_{\mu}^{i}\tau^{i} \Omega + i \frac{1}{2} g_1 \Omega \tau^3 a_{\mu}\\
D_{\mu}\psi^{a}_{L} =& \, \partial_{\mu}\psi^{a}_{L} - i g_2 \omega_{\mu}^{i}\tau^{i} \psi^{a}_{L} - i g_1 y_{a} a_{\mu} \psi^{a}_{L}\\
D_{\mu}\psi^{b}_{R} =& \, \partial_{\mu}\psi^{b}_{R} - i g_1 y_{b} a_{\mu} \psi^{b}_{R}\\
\end{split}
\end{equation}
the $SU(2)_{L}$ matrices are $\tau^{i}=\sigma^{i}/2$, and the $2\times 2$ matrix $\Omega$ is defined as
\begin{equation}
\Omega = \left(\begin{array}{cc}
\phi_1 & -\phi_2^{*}\\
\phi_2 & \phi_1^{*}
\end{array}\right)
\end{equation}
The gauge field-strength are
\begin{equation}
\begin{split}
\mathcal{F}_{\mu\nu}=& \partial_{\mu} a_{\nu}-\partial_{\nu} a_{\mu}\\
G_{\mu\nu}=& \partial_{\mu} \omega_{\nu}-\partial_{\nu} \omega - i g_2 \left[\omega_{\mu},\omega_{\nu}\right]\\
\end{split}
\end{equation}
This is exactly the Standard Model Lagrangian. However we assume $v^2$ much smaller than $\Lambda_2^2$, in order to have a confining $SU(2)_{L}$ gauge theory rather than spontaneously broken.

Before we introduce the Yukawa coupling and QCD, and if we set $g_1=0$, the Lagrangian has many global symmetries. We discuss the limit in which these effects are absent. In this case there is no distinction between quarks and leptons or between generations, and for the case $n_{f}=3$ the Lagrangian has a global $SU(12)$ symmetry, corresponding to a rotation of the left-handed $SU(2)_{L}$ doublet $\psi^{a}_{L}$ (we do not consider symmetries on the right-handed fermions $\psi^{b}_{R}$ since they do not feel the strong $SU(2)_{L}$ force). There is also a global $SU(2)$ symmetry (different from the gauged $SU(2)_{L}$) corresponding to the transformation
\begin{equation}
\Omega \rightarrow \Omega \mathcal{U}
\end{equation}
where $\mathcal{U}$ is a $SU(2)$ matrix. Thus the approximate global symmetry of the SCSM is $SU(12) \times SU(2)$. The symmetry breaking effects are electromagnetism, the Yukawa interactions and the strong interactions.

\subsection{Spectrum of the SCSM: `minimal sector'}
The physical states are $SU(2)_{L}$ singlets. The right-handed fermions and the $U(1)$ gauge boson $a_{\mu}$ are fundamental particles neutral under the $SU(2)_{L}$ interactions, and thus already $SU(2)_{L}$ singlets. The same is not true for fields interacting under the $SU(2)_{L}$ force, like the left-handed fermions and  the scalar $\phi$, which are bound into $SU(2)_{L}$ singlets.

In this model there is no electroweak symmetry breaking. The unbroken $U(1)$ is identified with the electromagnetic interactions, and its gauge coupling with the positron electric charge $e$. Thus we have $g_1=e$, and the $U(1)$ quantum numbers of physical particles must match the electric charge of the correspondent fields in the SM. We can use the global $SU(2)$ discussed above to classify these composite states.

We identify the left-handed physical fermions as the fermion-scalar bound states
\begin{equation}
F^{a}_{L} \doteq \Omega^{\dag}\psi^{a}_{L}=\left(\begin{array}{c}
\phi^{\alpha\,*}\psi^{a}_{L\,\alpha}\\
\phi_{\alpha}\epsilon^{\alpha\beta}\psi^{a}_{L\,\beta}
\end{array}\right)
\end{equation}
which transforms as a doublet under the global $SU(2)$. The symbol $\doteq$ means that the operator $\Omega^{\dag}\psi^{a}_{L}$ creates the state $F^{a}_{L} $ from the vacuum. The hypercharge of each composite fermion is the sum of the constituents hypercharges, which results in $\tau_3 + y_{a}$, or equivalently in the electric charge of the composite fermion. This is a check that the identification of the unbroken $U(1)$ in the SCSM with the electromagnetism in the SM gives the correct result. The $W^{i}$ gauge bosons of SM can be identified as the scalar-scalar bound states
\begin{equation}
W^{i}_{\mu} \doteq Tr\left[\Omega^{\dag}D_{\mu}\Omega \tau^{i}\right]
\end{equation}
which is a triplet under the global $SU(2)$. Finally we can make one other scalar-scalar bound state
\begin{equation}
H \doteq \frac{1}{2}Tr\left[\Omega^{\dag}\Omega\right]
\end{equation}
which is a $SU(2)$ singlet and corresponds to the SM Higgs boson. These are all the physical fields of the SCSM that we need to reproduce the ordinary weak-coupling SM, and they are listed in table \ref{tableSTATES}. If the model is correct these will not be the only physical states of the theory, and a rich spectrum of composite particles should emerge at the scale $\Lambda_2$. We will refer to the states found above as the `minimal sector' of the SCSM, whereas all the other resonances are called the `exotic sector'.

\begin{table}
\begin{center}
\begin{tabular}{|c|c|c|}\hline
\hline
state & operator & $SU(2)$ \\
\hline
$H$ & $\frac{1}{2}Tr\left[\Omega^{\dag}\Omega\right]$ & $\mathbf{1}$\\
$F^{a}_{L}$ & $\Omega^{\dag}\psi^{a}_{L}$ & $\mathbf{2}$\\
$W^{i}_{\mu}$ & $Tr\left[\Omega^{\dag}D_{\mu}\Omega \tau^{i}\right]$ & $\mathbf{3}$ \\
\hline
\hline
\end{tabular}
\end{center}
\caption{Composite physical states in the `minimal sector': the operator creates the correspondent state from the vacuum. The global $SU(2)$ representation is also given.}
\label{tableSTATES}
\end{table}

\subsection{Effective Lagrangian for the `minimal sector'}
Given the physical spectrum of our theory we can write the low-energy effective Lagrangian consistent with the global symmetries. Operators with dimension greater than four are suppressed by the scale $\Lambda_2$, and we can neglect them as long as we consider energy scales not above the $Z$ mass. The most general effective Lagrangian consistent with the global $SU(12) \times SU(2)$ and involving only the composite fermions and the composite vector-bosons is
\begin{equation}
\mathcal{L}_{eff}^{min,\,0} = i \bar{F}_{La}\,\slashed{\partial} F_{La} -\frac{1}{4} \mathbf{W}_{\mu\nu} \cdot \mathbf{W}^{\mu\nu} + \frac{1}{2} m_{W}^2 \mathbf{W}_{\mu}\mathbf{W}^{\mu} + g \mathbf{W}^{\mu} \mathbf{J}_{L\,\mu} + \ldots
\end{equation}
where $\mathbf{W}_{\mu\nu}=\partial_{\mu}\mathbf{W}_{\nu}-\partial_{\nu}\mathbf{W}_{\mu}$, $\mathbf{J}_{L}^{\mu}=\bar{F}_{La}\mathbf{\tau} \gamma^{\mu} F_{La}$ and $g$ is an effective coupling. The $\ldots$ stands for cubic and quartic vector-boson interactions.

The inclusion of the $U(1)$ gauge group breaks the $SU(2)$ global symmetry, since it allows the $W^3$ to mix with the $U(1)$ gauge boson. The resulting mass eigenstates are a massless state $A_{\mu}$ which we identify with the physical photon of the SM, and a massive state heavier than $W^{\pm}$ which we identify with the physical $Z$ boson of the SM. The low-energy effective Lagrangian including also the $U(1)$ gauge group and the right-handed fermions results in
\begin{equation}
\mathcal{L}^{min}_{eff} = \mathcal{L}_{eff}^{min,\,0} + i \bar{\psi}_{R b}\,\slashed{\partial} \psi_{R b} + e a_{\mu} J^{\mu}_{em} - \frac{1}{4} \mathcal{F}_{\mu\nu}\mathcal{F}^{\mu\nu} - \frac{k}{2} \mathcal{F}^{\mu\nu} W_{\mu\nu}^3 + \ldots
\end{equation}
where $a_{\mu}$ is the $U(1)$ gauge boson, $\mathcal{F}_{\mu\nu}$ the correspondent field strength and $J^{\mu}_{em}$ the electromagnetic current. We have dropped again any cubic or quartic vector-boson interactions.

The effective Lagrangian we have so far looks still very different from the SM Lagrangian, but we should reexpress it in terms of the mass eigenstates. The only non diagonal piece is the quadratic part of the vector-bosons, and the diagonalization is performed in the appendix \ref{appdiag}. The transformation which diagonalizes the Lagrangian is
\begin{equation}
\begin{split}
A_{\mu} =& k \,W^3_{\mu} + a_{\mu}\\
Z_{\mu} =& \sqrt{1-k^2} \,W_{\mu}^3
\end{split}
\end{equation}
The mass of the $Z$ boson results in $m_{Z}=m_{W}/\sqrt{1-k^2}$, and therefore we have the relation
\begin{equation}
\frac{m_{W}^2}{m_{Z}^2}= 1-k^2
\end{equation}
The neutral current Lagrangian is
\begin{equation}
\mathcal{L}^{min}_{NC} = e A_{\mu} J^{\mu}_{em} + Z_{\mu}\frac{g}{\sqrt{1-k^2}} \left(J_{L}^{3 \mu} -\frac{ek}{g} J^{\mu}_{em}\right)
\end{equation}
To reproduce the SM Lagrangian we have to impose the relations $k^2 = \sin^2\theta_{w}$ and $\frac{ek}{g}=\sin^2\theta_{w}$, where $\theta_{w}$ is the weak mixing angle in the SM. In order to satisfy these two relations we need the condition $k=e/g$. In the next section we explain why such relation is expected to be valid in the SCSM up to a very good approximation.

The physical fermions so far are massless. The way they get the mass in this model is by the inclusion of the Yukawa couplings $\lambda_{ab} \bar{\psi}^{a}_{L} \, \Omega \, \psi^{b}_{R}$ in the Lagrangian (\ref{SCSMLag}), which induces in the low-energy effective theory the interactions $H \bar{F}^{a}_{L} \psi^{b}_{R}$. The generated mass for the fermions results in $m_{ab} \simeq \lambda_{ab} \Lambda_2$. Therefore the composite scalar $H$ couples to the fermion-antifermion pairs with a strength proportional to the fermion mass, as the fundamental Higgs boson does in the Standard Model.

\subsection{Vector meson dominance}
The composite particles of the SCSM have nontrivial form factors characterized by the scale $\Lambda_2$. As an example we consider the electromagnetic form factor of the composite fermions. For the composite fermion-photon interaction there are two diagrams to lowest order: the direct coupling with the $U(1)$ gauge boson and the diagram where the $U(1)$ boson goes first into a $W^3$, and then the $W^3$ couples to the composite fermions. The resulting amplitude is
\begin{equation}
e \langle \bar{F}_{La} F_{La}| J^{\mu}_{em} | 0 \rangle = e \bar{U}_{L} \gamma^{\mu} Q^{a} V_{L} - kg\frac{q^2}{q^2-m_{W}^2}\bar{U}_{L} \gamma^{\mu}\tau^3 V_{L}
\end{equation}
Using the relation $Q^a=y^a+\tau^3$ we can decompose the current into singlet and vector pieces (with respect to the $SU(2)$ global symmetry)
\begin{equation}
e \langle \bar{F}_{La} F_{La}| J^{\mu}_{em} | 0 \rangle = e \bar{U}_{L} \gamma^{\mu} y^{a} V_{L} +e\left[1- \frac{kg}{e}\frac{q^2}{q^2-m_{W}^2}\right]\bar{U}_{L} \gamma^{\mu} \tau^3 V_{L}
\end{equation}
and we can get the vector form factor
\begin{equation}
F_{V}(q^2) = 1- \frac{kg}{e}\frac{q^2}{q^2-m_{W}^2}
\end{equation}
Since the interaction which binds the composite fermions is asymptotically free the electromagnetic form factors should vanish for $q^2 \rightarrow \infty$. By imposing this condition on the vector form factor found above we get the relation
\begin{equation}
k=\frac{e}{g}
\end{equation}
which is what we need to reproduce the SM phenomenology. The singlet form factor is constant and equal to one, and for the same reason it has to vanish for $q^2 \rightarrow \infty$. Hence we expect the presence of iso-singlet excited states which make it vanish. It is also possible to show that if we impose the same condition for the charge and magnetic moment form factors and two-photon couplings of the $W$ we can get the same quartic and cubic interaction of the SM \cite{Claudson:1986ch}.

We do not expect the relation $k=e/g$ to be exact, since intermediate excited states can give an additional contribution to $F_{V}(q^2)$ and then modify that relation. The assumption that the vector form factor is saturated only by the $W$ boson is known as `vector meson dominance' (VMD), from its use in hadronic physics. Since the SM reproduces the observed data with extremely good accuracy we expect that the deviations from this relation are small, and therefore we expect the resonant states to be heavy and weakly coupled. For this reason the relevant corrections given by the excited states are only at tree-level.

\section{Adding an excited triplet}
\label{triplet}
The resonant states we consider are the first radial excitations of the vector-boson triplet $W^{i}$ introduced previously. We call these lowest excitations $W^{\prime i}$. The neutral component $W^{\prime 3}$ can mix with the photon also in this case, and the resulting mass eigenstate is denoted by $Z^{\prime}$, which is again heavier than the charged component $W^{\prime \pm}$. For the reason explained before we consider the inclusion of this triplet as a perturbation of the SCSM `minimal sector`.

The full effective Lagrangian has the following form
\begin{equation}
\mathcal{L}_{eff} = \mathcal{L}_{eff}^{min} + \mathcal{L}^{W^{\prime}}_{eff}
\end{equation}
where $\mathcal{L}_{eff}^{min}$ is the effective Lagrangian for the minimal sector of the SCSM found in the previous section and $\mathcal{L}^{W^{\prime}}_{eff}$ contains the new operators due to the $W^{\prime}$ triplet. The most general Lagrangian we can write with the new fields is
\begin{equation}
\mathcal{L}^{W^{\prime}}_{eff} = -\frac{1}{4} \mathbf{W^{\prime}}_{\mu\nu} \cdot \mathbf{W^{\prime}}^{\mu\nu} + \frac{1}{2} m_{W^{\prime}}^2 \mathbf{W^{\prime}}_{\mu}\mathbf{W^{\prime}}^{\mu} + g^{\prime} \mathbf{W}^{\prime \mu} \mathbf{J}_{L\,\mu} - \frac{k^{\prime}}{2} \mathcal{F}^{\mu\nu} W^{\prime 3}_{\mu\nu}  + \ldots
\end{equation}
where we again do not consider cubic and quartic terms. There is no mixing between $W$ and $W^{\prime}$ because we have diagonalized the Lagrangian before the inclusion of the $U(1)$ gauge group, which causes the mixing between the $W^{\prime 3}$ and the $U(1)$ gauge boson. Thus adding a vector isotriplet adds three free parameters to the effective Lagrangian. By a redefinition of the fields it is possible to show that only the relative sign between $k^{\prime}$ and $g^{\prime}$ is physical, therefore we restrict to positive values for $k^{\prime}$. We expect the excited states heavy and weakly coupled. We define
\begin{equation}
\epsilon=\frac{k^{\prime}}{k}  \;\;\;\;\;\;\;\;\;\;\;\;  \lambda=\frac{g^{\prime}}{g} \;\;\;\;\;\;\;\;\;\;\;\; \mu= \frac{m_{W}^2}{m_{W^{\prime}}^2}
\label{small}
\end{equation}
We perform the analysis for small couplings: $\epsilon\ll 1$ and $\lambda\ll 1$. We do not need any assumption about $\mu$ in our computations.

The presence of these new states modifies the vector electromagnetic form factor, and therefore violates the VMD hypotesis, since now the form factor is not saturated by only the $W^{3}$ vector-boson anymore. To see this more explicitly we can compute again the vector form factor and we find an extra term
\begin{equation}
F_{V}(q^2) = 1- \frac{kg}{e}\frac{q^2}{q^2-m_{W}^2}- \frac{k^{\prime}g^{\prime}}{e}\frac{q^2}{q^2-m_{W^{\prime}}^2} = 1 - \frac{kg}{e} \left[\frac{q^2}{q^2-m_{W}^2} + \epsilon\lambda\frac{q^2}{q^2-m_{W^{\prime}}^2}\right]
\end{equation}
The new contribution corresponds to the process when the $U(1)$ boson goes first into a $W^{\prime 3}$, which then couples to the composite fermions. By imposing that this form factor vanishes for infinite momentum transfer we find the condition
\begin{equation}
e = kg \left(1+\epsilon\lambda\right)
\label{VMDviol}
\end{equation}
This explain why it is relevant to consider these excitations: they correspond to relaxing the VMD hypothesis, and the amount of VMD violation is $\epsilon\lambda$.

The quadratic part of the Lagrangian when we include the excited triplet is nondiagonal. The charged states are already diagonal and canonically normalized, whereas for the neutral states we have a situation similar to the minimal sector alone, with the additional complication that we have now mixing among three neutral fields. The diagonalization now is more involved and it is performed in the appendix \ref{appdiag}.
The result is the following
\begin{equation}
\begin{split}
a_{\mu} = & A_{\mu} - k W^{3}_{\mu} -
k^{\prime} W^{\prime 3}_{\mu}\\
W^{3}_{\mu}= & \alpha_{1} Z_{\mu} + \alpha_{2} Z_{\mu}^{\prime}\\
W^{\prime 3}_{\mu}=& \alpha_{3} Z_{\mu} + \alpha_{4} Z_{\mu}^{\prime}
\label{cbasis}
\end{split}
\end{equation}
where the coefficients $\alpha_{i}$ are functions of the parameters $(k, \epsilon, \mu)$. Since we are considering the inclusion of this excited triplet as a perturbation of the minimal sector of the SCSM we expect a small deviation from the zero-th order solution, corresponding to the case of no excited triplet. Hence it is natural to expand around this solution
\begin{equation}
\alpha_1 = \frac{1}{c_{w}} + \delta_1 \hspace{2cm} \alpha_2 = \delta_2 \hspace{2cm} \alpha_3 = \delta_3 \hspace{2cm} \alpha_4 = 1 + \delta_4
\end{equation}
where we rename $k=s_{w}$ because when the excited triplet is absent the parameter $k$ has the same role as the sine of the weak mixing angle in the conventional SM, as explained before. We define also the cosine of the weak mixing angle as $c_{w}^2=\sqrt{1-s_{w}^2}$, and we finally express the change of basis as
\begin{equation}
\begin{split}
a_{\mu} = & A_{\mu} - s_{w}\left(\frac{1}{c_{w}}+\delta_1+\epsilon \delta_3\right) Z_{\mu} - s_{w}\left(\delta_2+\epsilon + \epsilon\delta_4\right) Z^{\prime}_{\mu} \\
W^{3}_{\mu}= & \left(\frac{1}{c_{w}} + \delta_1\right) Z_{\mu} + \delta_{2} Z_{\mu}^{\prime}\\
W^{\prime 3}_{\mu}=& \delta_{3} Z_{\mu} + \left(1 + \delta_4\right) Z_{\mu}^{\prime}
\end{split}
\label{cb}
\end{equation}
The coefficients $\delta_{i}$ are derived in the appendix \ref{appdiag}.

The masses of the neutral vector-bosons are the eigenvalues of the mass matrix derived in the appendix \ref{appdiag}, and they results in
\begin{equation}
m_{A}^2 = 0, \hspace{0.6cm} m_{Z,Z^{\prime}}^2 = \frac{1-s_{w}^2+\mu-\epsilon^2 s_{w}^2 \mu \mp \sqrt{\left(1-s_{w}^2+\mu-\epsilon^2\mu s_{w}^2\right)^2-4 \mu \left(1-s_{w}^2-\epsilon^2 s_{w}^2\right)}}{2\mu\left(1-s_{w}^2-\epsilon^2 s_{w}^2\right)} m_{W}^2
\label{Zmass}
\end{equation}
It is possible to check that in the limit $\epsilon \rightarrow 0$ and $\mu \rightarrow 0$ we have the relation $m_{Z} c_{w} = m_{W}$, which is the correct Standard Model limit.

\section{The low-energy effective Lagrangian}
\label{efflag}
In this section we derive the effective Lagrangian for the theory below the mass scale of the heavy vector-bosons $m_{W^{\prime}}$. In the following section we will use this Lagrangian to compute the contribution to the electroweak observables given by the excited states. In the process of integrating out the heavy degrees of freedom we leave in the spectrum of our low-energy theory only the particles which have been detected so far, therefore we integrate out the Higgs boson and the heavy vector-bosons $W^{\prime \pm}$ and $Z^{\prime}$. The virtual effects of the Higgs boson are the same as in the conventional SM\footnote{In the SCSM there might be some extra couplings of the composite Higgs boson with the vector bosons \cite{Skiba:1995ep}. In this work we assume a Higgs sector as in the conventional SM.}, hence we have only to take care of the heavy vector-bosons.

In our model in principle we have 7 parameters: $e$, $k=s_{w}$, $g$, $m_{W}$, $k^{\prime}$, $g^{\prime}$, $m_{W^{\prime}}$. First of all we trade the three new parameters for the ratios with the correspondent parameters of the minimal sector: $\epsilon$, $\lambda$ and $\mu$, as they are defined in equation (\ref{small}). In addition we have that the other four parameters are not independent: a relation among them is given by the equation (\ref{VMDviol}). We take this equation as an expression for $g$. We finally trade $m_{W}$ for $m_{Z}$, since we are going to use the latter as an input parameters when we make the electroweak fit. In conclusion we have 6 parameters: $e$, $s_{w}$ and $m_{Z}$ already present when we consider the minimal sector, $\epsilon$, $\lambda$ and $\mu$ from the excited states. The effective Lagrangian we are going to construct has to be a function of only these 6 parameters. In the remaining part of this section we list all the operators of the low-energy effective Lagrangian which are relevant for the electroweak analysis.

\subsection{Vector-bosons kinetic and mass terms}
After the diagonalization we have a low-energy effective Lagrangian for the vector-bosons
\begin{equation}
\mathcal{L}_{vector} = -\frac{1}{4} A_{\mu\nu}A^{\mu\nu} - \frac{1}{4} Z_{\mu\nu}Z^{\mu\nu} -\frac{1}{2} W_{\mu\nu}^{\dag} W^{\mu\nu} + m_{W}^2 W_{\mu}^{\dag} W^{\mu} + \frac{1}{2} m_{Z}^2 Z_{\mu} Z^{\mu}
\end{equation}
If $m_{W}=m_{Z} c_{w}$ then this would just be the SM Lagrangian for the vector bosons. The relation between the two masses is now modified by the mixing effects, and since we are using $m_{Z}$ as a parameter of the Lagrangian we have to compute the correction to $m_{W}$. The relation between $m_{Z}$ and $m_{W}$ is given in equation (\ref{Zmass}). Since we are considering the inclusion of the triplet as a perturbation we can expand this relation for small values of the parameter $\epsilon$, and up to quadratic terms we get
\begin{equation}
m_{W}^2 = m_{Z}^2 \left[c_{w}^2+\frac{s_{w}^4}{c_{w}^2-\mu}\mu\epsilon^2\right]
\end{equation}

\subsection{Neutral current interactions: $Z$ boson vertex}
The diagonalization performed previously changes the expression for the neutral current (NC) interactions, which in the original basis is
\begin{equation}
\mathcal{L}_{NC} = e J^{\mu}_{em} a_{\mu} + g J^{3 \mu}_{L} W^{3}_{\mu} + g^{\prime} J^{3 \mu}_{L} W^{\prime 3}_{\mu}
\label{NCL}
\end{equation}
where the left and electromagnetic currents are respectively
\begin{equation}
J^{3 \mu}_{L}=\bar{F}_{La} \gamma^{\mu} \tau^3 F_{La} \hspace{2cm}  J^{\mu}_{em}= \bar{F}_{La} \gamma^{\mu} \left(\tau^3+y_{a}\right)F_{La}+y_{b}\bar{F}_{Rb} \gamma^{\mu} F_{Rb}
\end{equation}
The right-handed physical fermions $F_{Rb}$ are the point-like fermions $\psi_{Rb}$ of the SCSM fundamental Lagrangian, each one with hypercharge $y_{b}$. We plug the expressions in equation (\ref{cb}) into the Lagrangian in equation (\ref{NCL}), in order to have only the mass eigenstates. Since we want to compare our model with experiments performed at the $Z$ pole we are interested only in the interaction
\begin{equation}
\mathcal{L}_{NC, Z} = \frac{1}{c_{w}} \left[g \left(1+\delta_1 c_{w} + \lambda \delta_3 c_{w}\right) J^{3 \mu}_{L} - e s_{w} \left(1+\delta_1 c_{w}+ \epsilon \delta_3 c_{w}\right)J^{\mu}_{em}\right] Z_{\mu}
\end{equation}
From this expression it is manifest that if we turn the mixing off ($\epsilon=0$, and then $\delta_{i}=0$), we get the NC interactions of the minimal sector of the SCSM. The expression has to be a function of the free parameters listed above, thus we have to express $g$ as a function of these free parameters of the Lagrangian, as in equation (\ref{VMDviol}). The final result for these interactions is
\begin{equation}
\mathcal{L}_{NC, Z} = e J^{\mu}_{em} A_{\mu} + \frac{e}{s_{w} c_{w}} \left[\frac{1}{1+\epsilon\lambda} \left(1+\delta_1 c_{w} + \lambda \delta_3 c_{w}\right) J^{3 \mu}_{L} -  s^2_{w} \left(1+\delta_1 c_{w}+ \epsilon \delta_3 c_{w}\right)J^{\mu}_{em}\right] Z_{\mu}
\label{NClag}
\end{equation}

\section{Contribution to the ElectroWeak Observables}
\label{EWobs}
In the electroweak fit we choose $\alpha(m_{Z})$, $G_{F}$ and $m_{Z}$ as input parameters, since they are the best measured parameters of the SM electroweak sector. In our model we have 6 parameters, three already present in the minimal sector of the SCSM, three from the excited states. We trade the first three with the three best measured electroweak observables, and then we have a prediction for all the electroweak observables as a function of the three parameters of the excited triplet $\epsilon$, $\lambda$ and $\mu$. Two of our parameters are already input parameters ($m_{Z}$ and $e$, which is related to $\alpha$). We have to find a way to relate the bare parameter $s_{w}$ to our third input parameter, the Fermi constant. We define as usual the weak mixing angle $s_{0}$ from the $Z$ pole \cite{Peskin:1990zt,Peskin:1991sw} \footnote{We define also $c_0 = \sqrt{1-s_0^2}$.}
\begin{equation}
\frac{G_{F}}{\sqrt{2}} = \frac{e^2}{8 \,s_{0}^2 c_{0}^2\,m_{Z}^2}
\end{equation}
If we compare this expression with $G_{F}$ computed in our model we have a relation between the bare parameter $s_{w}$ and the input parameter $s_{0}$. The Fermi constant is measured in muon decay, therefore
\begin{equation}
\frac{G_{F}}{\sqrt{2}} = \left(\frac{g^2}{8 m_{W}^2}+\frac{g\prime^2}{8 m_{W^{\prime}}^2}\right) = \frac{e^2}{8 s_{w}^2 c_{w}^2 m_{Z}^2 } \frac{\left(1 + \mu \lambda^2\right)}{\left(1+\epsilon\lambda\right)^2\left(1+\frac{s_{w}^4}{c_{w}^2\left(c_{w}^2-\mu\right)}\mu\epsilon^2\right)}
\end{equation}
If we now compare the two expressions for $G_{F}$ we get
\begin{equation}
s_{0}^2 c_{0}^2 = s_{w}^2 c_{w}^2 \frac{\left(1+\epsilon\lambda\right)^2}{\left(1 + \mu \lambda^2\right)} \left(1+\frac{s_{w}^4}{c_{w}^2\left(c_{w}^2-\mu\right)}\mu\epsilon^2\right)
\label{sw2}
\end{equation}
and we can finally get an expression for $s_{w}$ as a function of the input parameters and the three new parameters describing the excited states.

In the global fit we consider observables measured in $e^{+}e^{-}$ colliders at the $Z$ pole, and we follow Burgess et al. \cite{Burgess:1993vc} in parameterizing the deviation of the fermion couplings with the Z boson
\begin{equation}
\mathcal{L}^{Z}_{NC} = \frac{e}{s_{w} c_{w}} \sum_{f} \bar{f} \gamma^{\mu} \left[g_{L}^{f} P_{L} + g_{R}^{f} P_{R}\right] f Z_{\mu}
\label{LNCZ}
\end{equation}
where $P_{L,R}$ are left-right projector and the couplings are written as
\begin{equation}
\begin{array}{llllllllllll}
g_{L}^{f} = g_{L}^{f, SM} + \delta g_{L}^{f}, &  &  &  &  &  &  &  &  &  &  & g_{L}^{f, SM}=\tau_{3}^{f}-s_{w}^2 Q_{f}\\
g_{R}^{f} = g_{R}^{f, SM} + \delta g_{R}^{f}, &  &  &  &  &  &  &  &  &  &  & g_{R}^{f, SM}=-s_{w}^2 Q_{f}
\end{array}
\end{equation}
It is possible to compute the shifts $\delta g_{L,R}^{f}$ from the effective Lagrangian in equation (\ref{NClag}): we use the fact that $\epsilon$ and $\lambda$ are assumed to be small, then we expand up to quadratic terms (the $\epsilon$ dependence for $\delta_{i}$ is given in the appendix \ref{appdiag}), and we get
\begin{equation}
\begin{split}
\delta g_{L}^{f} = & \tau_{3}^{f}\left(\delta_1 c_{w} + \lambda \delta_3 c_{w} -\epsilon\lambda \right) - s^2_{w} Q_{f} \left(\delta_1 c_{w}+ \epsilon \delta_3 c_{w}\right)\\
\delta g_{R}^{f} = & - s^2_{w} Q_{f} \left(\delta_1 c_{w}+ \epsilon \delta_3 c_{w}\right)
\end{split}
\end{equation}

The observables used in the global fit are listed in the appendix \ref{appobs}, whereas their experimental values and the Standard Model predictions can be found in the appendix \ref{appGFITTER}.

\section{Results of the global fit}
We now take the expressions for the electroweak observables given in appendix \ref{appobs} and compare them with the experimental data. We define the $\chi^2$ function as
\begin{equation}
\chi^2\left[\epsilon, \mu, \lambda \right] = \sum_{i} \left(\frac{\mathcal{O}_{i}\left(\epsilon, \mu, \lambda\right) - \mathcal{O}_{i}^{\exp}}{\sigma_{i}}\right)^2
\end{equation}
The sum runs over all the observables listed in the table \ref{tableOBS}, where also the experimental results and the SM theoretical predictions are shown. We can find the 95\% CL region in the $\left(\epsilon, \mu, \lambda\right)$ three dimensional space, but before showing the results we have to discuss the limits coming from the negative collider searches, since no $W^{\prime}$ triplet has been observed yet. At LEP a $W^{\prime}$ could have been produced by a s-channel photon or $Z$ exchange, and the cross section is large enough to rule out a $W^{\prime}$ mass of half of the center of mass energy, $m_{W^{\prime}} \leq \sqrt{s}/2 \simeq 105 \, \rm{GeV}$ \cite{pdg}. The $W^{\prime}$ have been searched for also at Run II at the Tevatron, and the current best limit of the $W^{\prime}$ coupling to quarks as a function of the $W^{\prime}$ is obtained considering $\bar{t} b$ and $t \bar{b}$ in the final states \cite{cdf9150}. In particular CDF ruled out a $W^{\prime}$ with mass $300\,\rm{GeV}$ and coupling $|\lambda| \simeq 0.4$, whereas for higher values of the $W^{\prime}$ mass the limit on the coupling gets weaker. Such a result is obtained by assuming a $W^{\prime}$ coupling only to fermions, but in the SCSM, and above the threshold $m_{W^{\prime}} \geq m_{Z} + m_{W} \simeq 172 \, \rm{GeV}$, the dominant decay channel would be a $W Z$ in the final state. However to be safe with that limit we consider only the region of parameter space such that $|\lambda| \lesssim 0.4$.

\label{ResultsFit}
\begin{figure}
\begin{center}
$\begin{array}{c@{\hspace{1in}}c}
\epsfxsize=2.6in
\epsffile{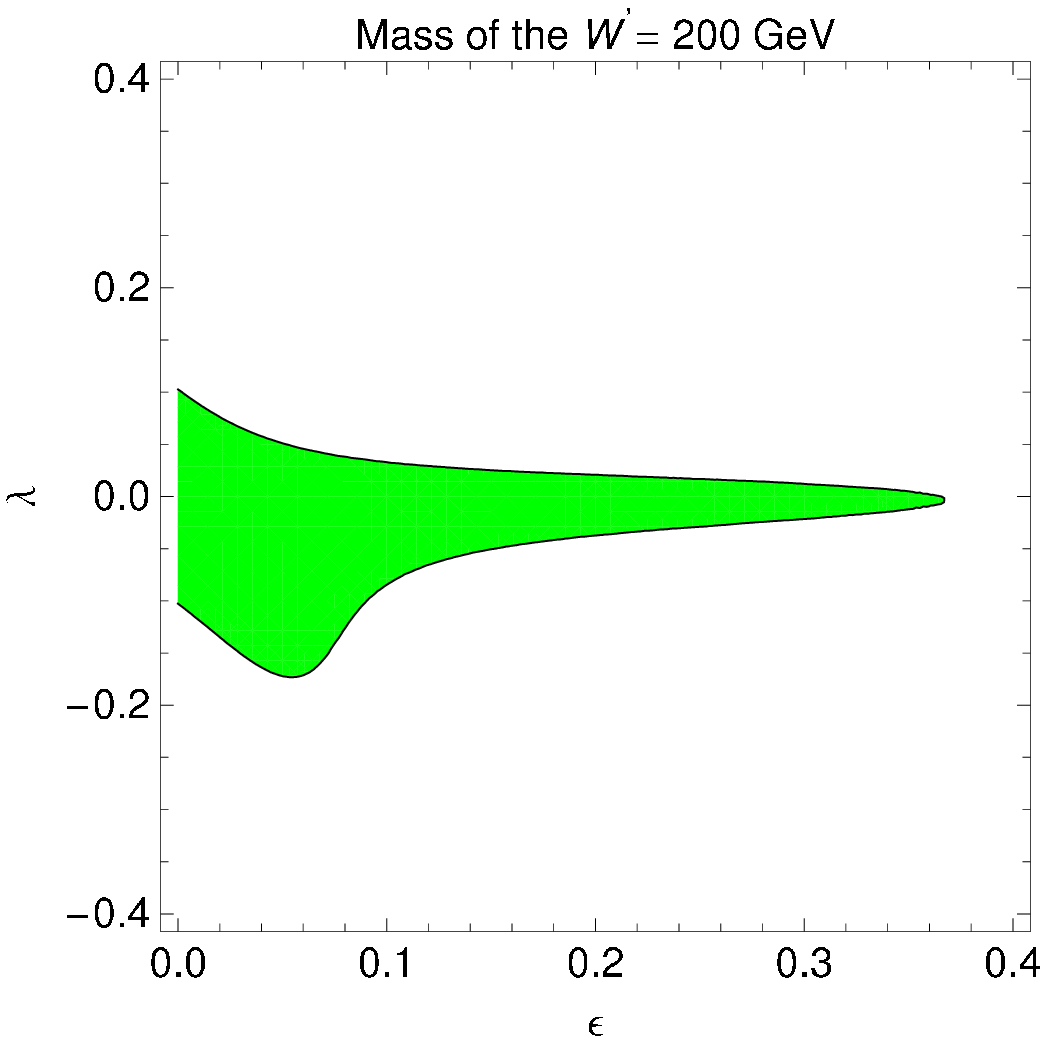} &
	\epsfxsize=2.6in
	\epsffile{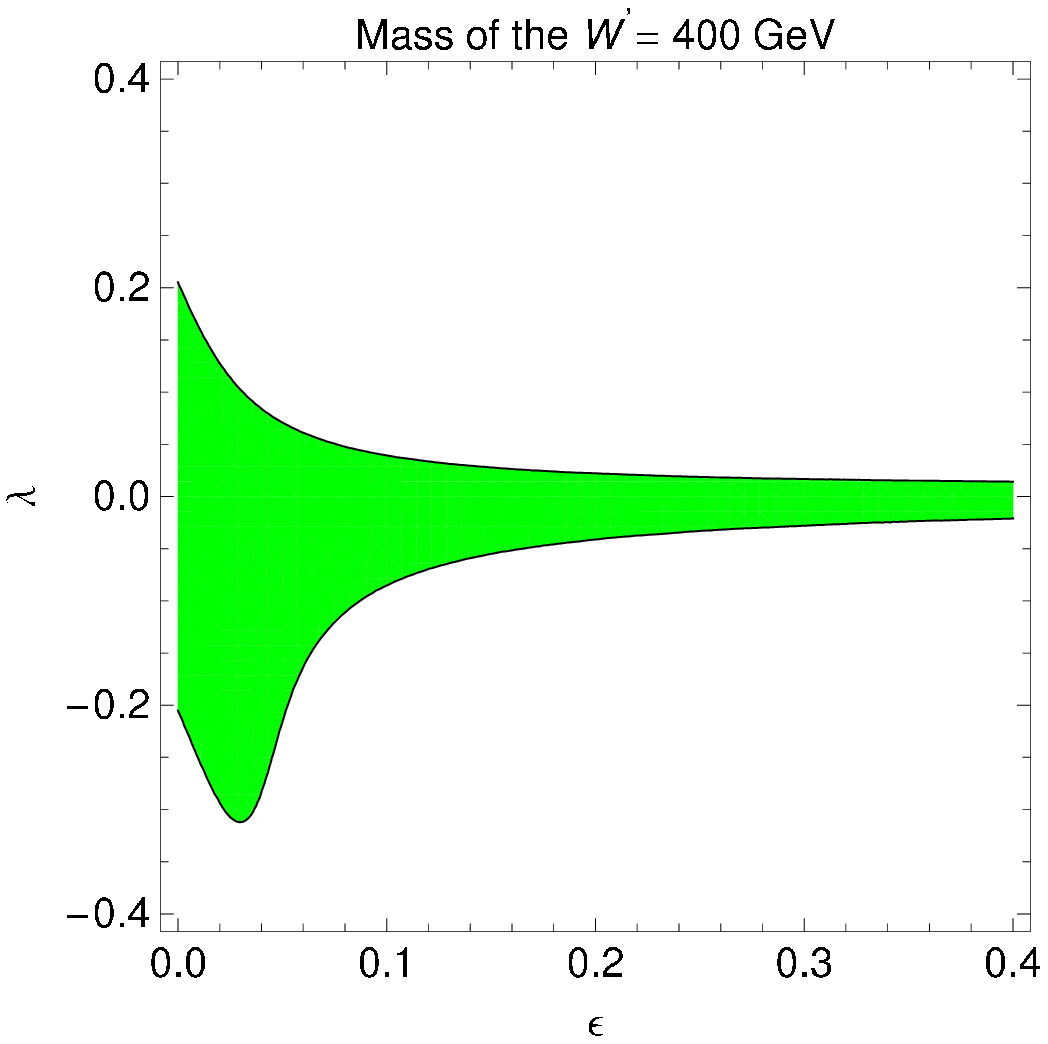} \\ [0cm]
\mbox{\bf (a)} & \mbox{\bf (b)} \\
\epsfxsize=2.6in
\epsffile{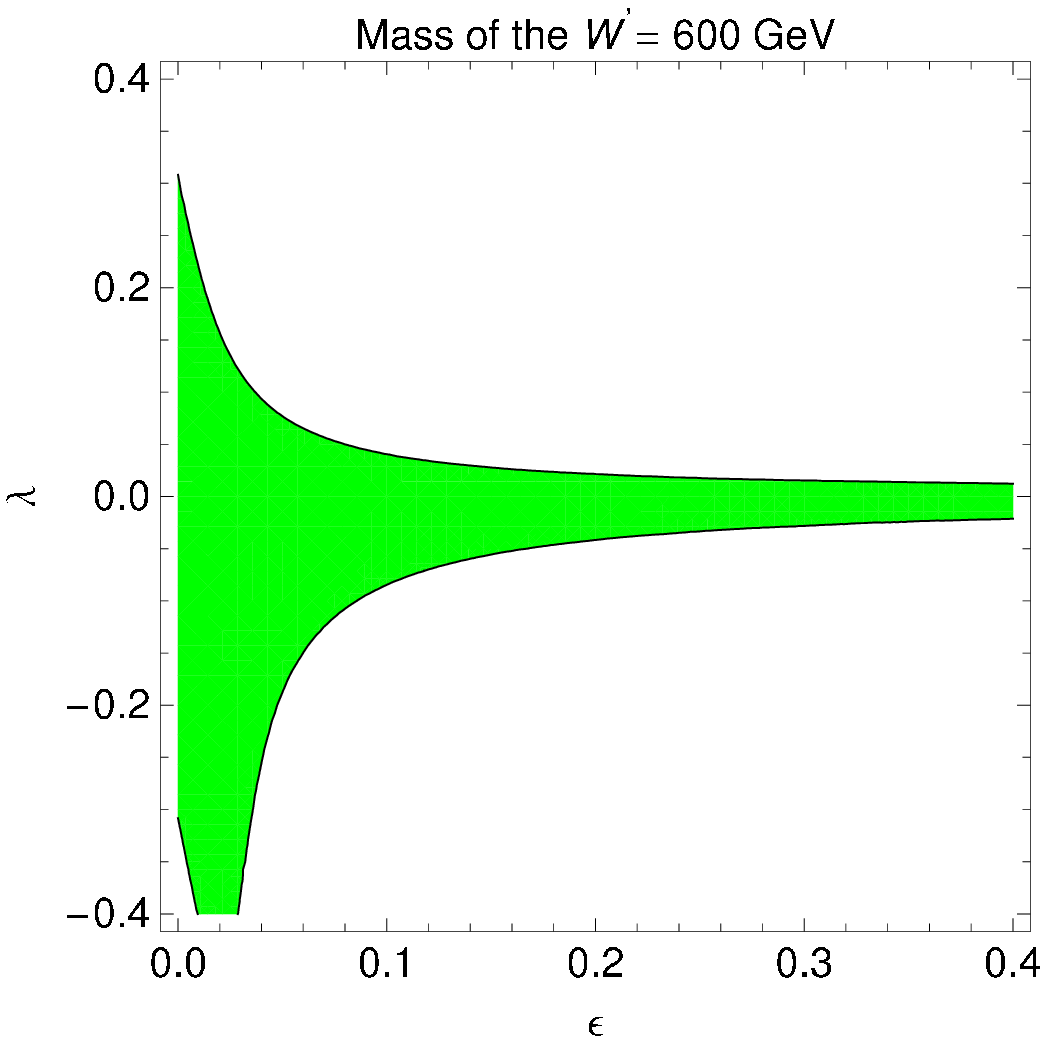} &
	\epsfxsize=2.6in
	\epsffile{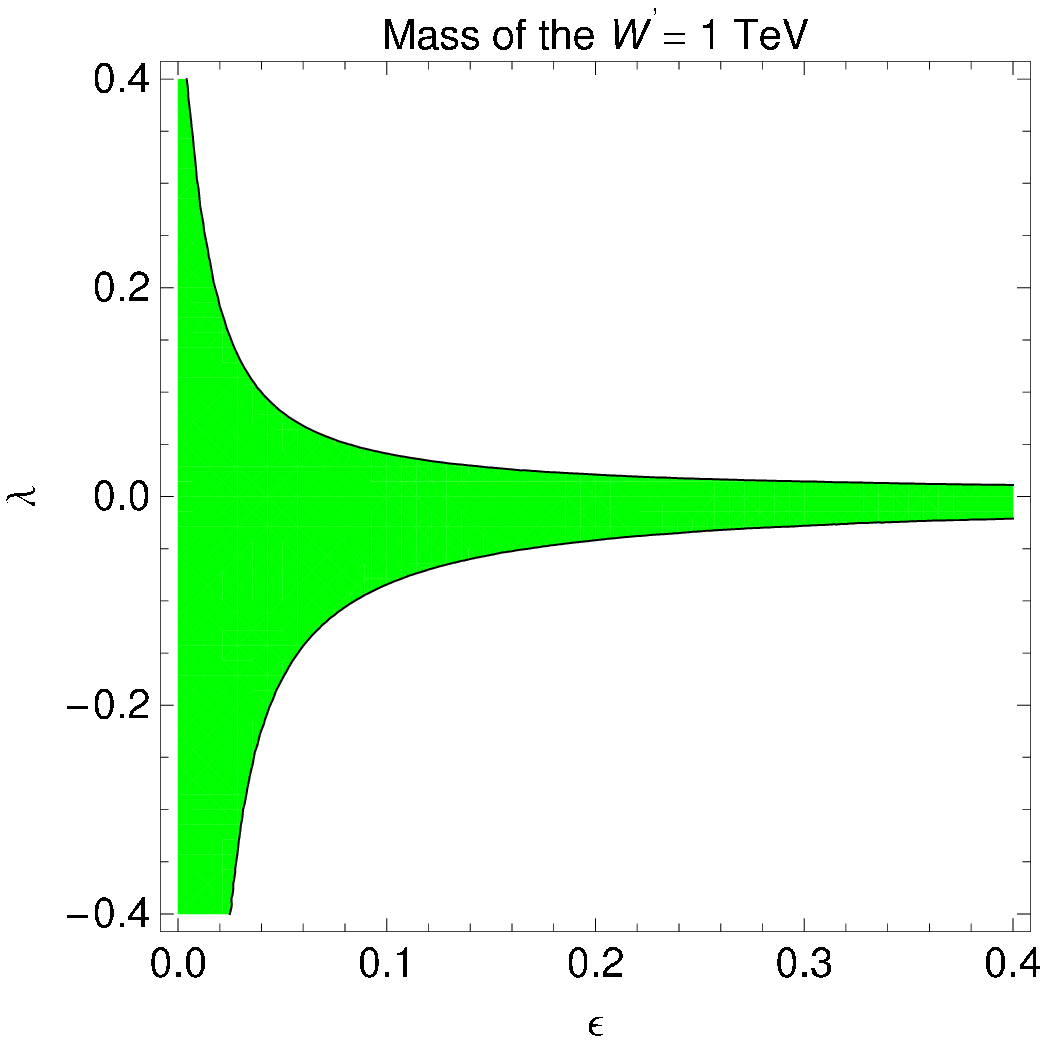} \\ [0.4cm]
\mbox{\bf (c)} & \mbox{\bf (d)}
\end{array}$
\end{center}
\caption{
Allowed 95\% CL region in the $\left(\epsilon, \lambda\right)$ plane for different values of the $W^{\prime}$ mass: (a) $m_{W^{\prime}}=200 \;\rm{GeV}$; (b) $m_{W^{\prime}}=400 \;\rm{GeV}$; (c) $m_{W^{\prime}}=600 \;\rm{GeV}$; (d) $m_{W^{\prime}}=1 \;\rm{TeV}$. Since only the relative sign between the two parameters is physical we plot only the region $\epsilon \geq 0$.
}
\label{3-2}
\end{figure}
We can now discuss the result of the global fit with the electroweak observables. First of all we find the 95\% CL region in the three dimensional $\left(\epsilon, \mu, \lambda\right)$ space. It is interesting to consider the allowed values for the product $\epsilon \lambda$, since this quantity is the amount of VMD violation. We find
\begin{equation}
-0.0084 \leq \epsilon \lambda \leq 0.0042 \hspace{3cm} 95\% \rm{CL}
\label{VMDviol95CL}
\end{equation}
Hence at 95\% CL the VDM hypothesis can be violated 1\% at most. This is equivalent to say that the electromagnetic vector form factor for fermions must be saturated by the $W$ boson at least by 99\%. We finally show sections of the $\left(\epsilon, \lambda\right)$ plane for fixed values of the $m_{W^{\prime}}$ mass, and we plot the 95\% CL allowed regions. The results are in figure \ref{3-2}, where we can see that in general the effective couplings are constrained to be rather small. We can also see that as we increase the $m_{W^{\prime}}$ mass the allowed parameter space becomes larger, as expected since the heavy triplet decouples from the low-energy physics. If $m_{W^{\prime}} \gtrsim 1\, \rm{TeV}$ the sections in the $\left(\epsilon, \lambda\right)$ plane are identical, because we get to the limit in equation (\ref{VMDviol95CL}). For such big values of the $W^{\prime}$ mass when one coupling is very small the other is not constrained. This is a consequence of the formulae for all the corrections to the electroweak observables, since whenever two of the three parameters $\left(\epsilon, \mu, \lambda\right)$ are vanishing all the corrections are also vanishing, leaving the other parameter completely unconstrained.

\section{Sensitivity to the Higgs mass}
\label{Higgs}
In our analysis we made two assumptions about the Higgs boson: we assumed a Higgs sector identical to the SM Higgs, and we also put by hand a value $115 \, \rm{GeV}$ for its mass. We do not expect that the first assumption changes the conclusions of our analysis, but a larger Higgs mass contributes appreciably to the electroweak observables through its virtual effects, and it may alter our results. In this last section we consider the case of a heavy Higgs boson. In the Standard Model the electroweak data prefer a light Higgs: the best fit value for the Higgs mass is $84 \, \rm{GeV}$, with a 95\% CL upper limit  of $154 \, \rm{GeV}$, and once we add the information that no Higgs has been found up to  $114.4 \, \rm{GeV}$ \cite{Barate:2003sz} this limit increases to $185 \, \rm{GeV}$\cite{Collaboration:2008ub}. However the EWPT can accomodate a heavy Higgs in a rather simple way, they just need new physics contributing to the electroweak observable and compensating the effect of a larger Higgs mass \cite{Peskin:2001rw}, as proposed in references \cite{Barbieri:2006dq,Barbieri:2006bg,D'Eramo:2007ga,Enberg:2007rp}.
\begin{figure}
\begin{center}
$\begin{array}{c@{\hspace{1in}}c}
\epsfxsize=2.6in
\epsffile{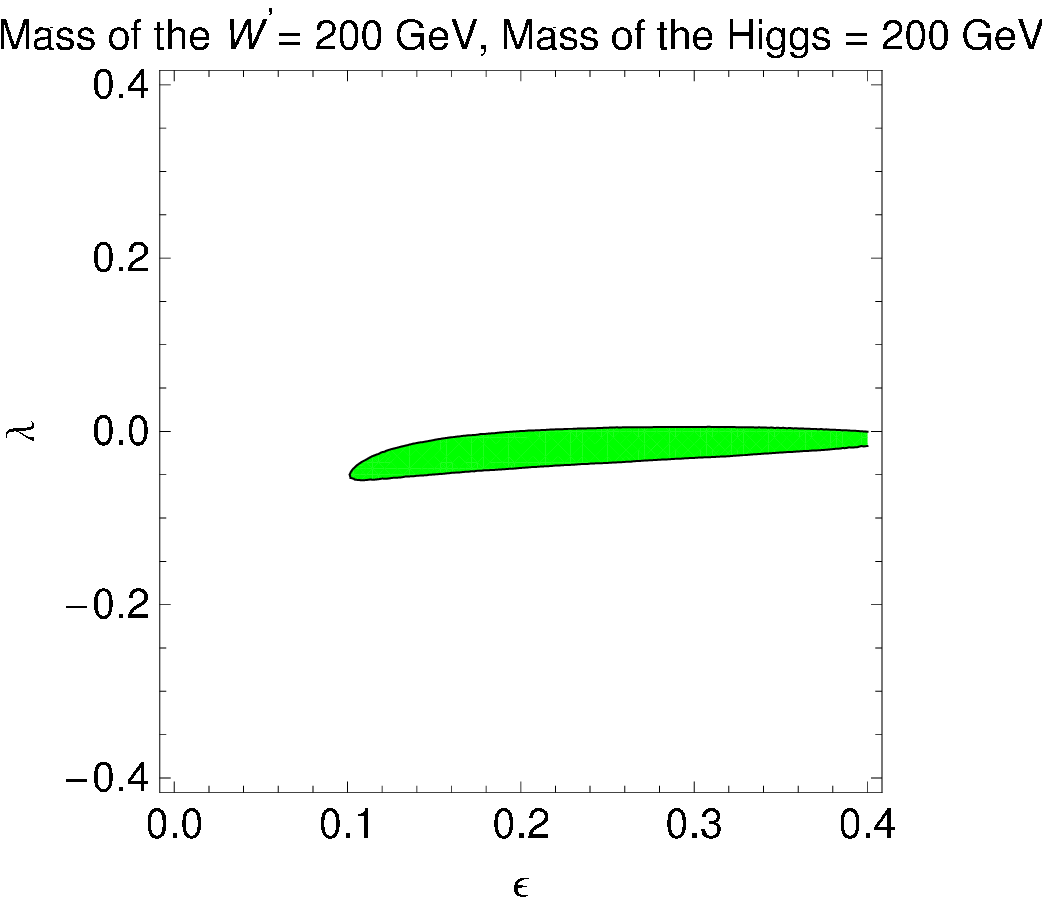} &
	\epsfxsize=2.6in
	\epsffile{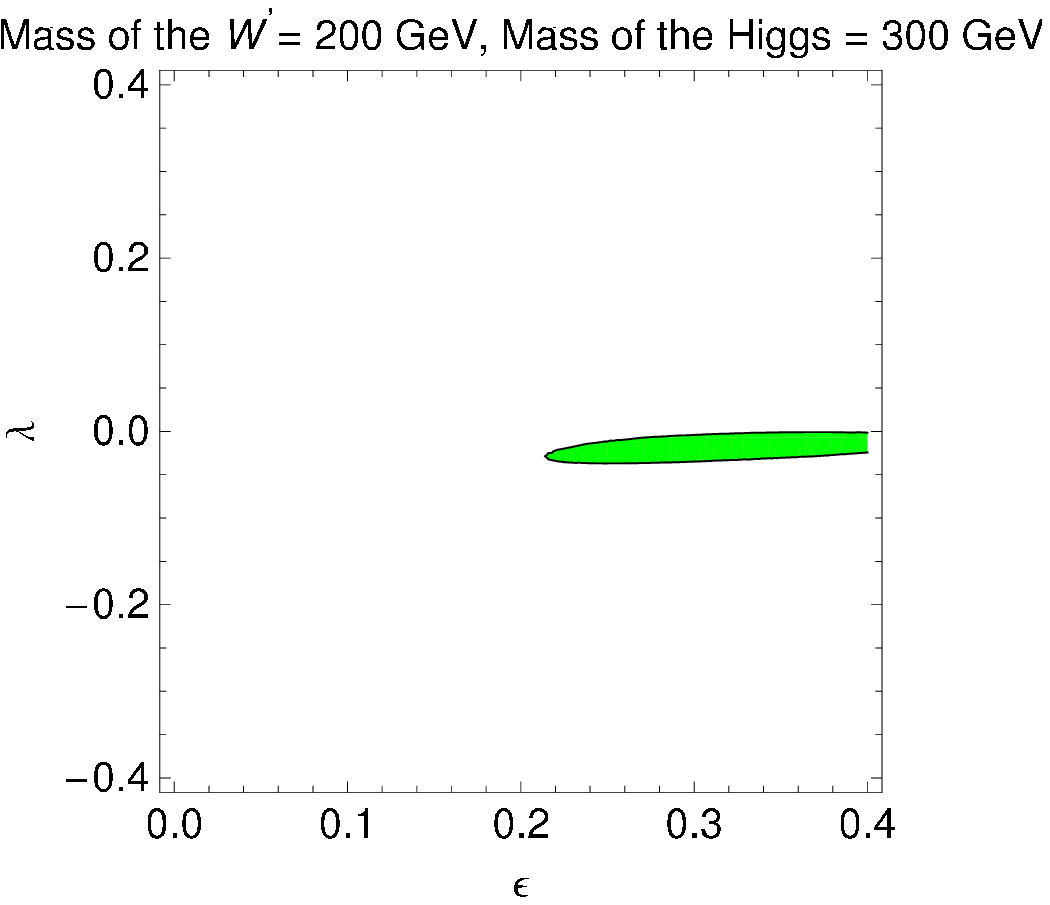} \\ [0cm]
\mbox{\bf (a)} & \mbox{\bf (b)}
\end{array}$
\end{center}
\caption{
Allowed 95\% CL region in the $\left(\epsilon, \lambda\right)$ plane for $W^{\prime} = 200 \;\rm{GeV}$ and different values of the Higgs mass: (a) $m_{h}=200 \;\rm{GeV}$; (b) $m_{h}=300 \;\rm{GeV}$.}
\label{heavyhigfig}
\end{figure}

In our case the new physics might be the $W^{\prime}$ triplet, which contributes to the electroweak observables as shown previously. We make again the same electroweak global fit as done before, but this time the Standard Model input values are computed for different values of the Higgs mass. We consider four cases: $m_h = 200 \,\rm{GeV}$, $300 \,\rm{GeV}$, $400 \,\rm{GeV}$, $500 \,\rm{GeV}$. A heavy Higgs boson up to $m_{h} \simeq 300 \, \rm{GeV}$ is allowed by the EWPT only for relatively low values of the $W^{\prime}$ mass, of the order of $m_{W^{\prime}} \simeq 200 \, \rm{GeV}$, as shown in the figure \ref{heavyhigfig}. However the allowed 95\% CL region is really tiny, much worse than for a light Higgs boson. If we further increase the Higgs boson mass there is no allowed region at 95\% CL for $m_{W^{\prime}} \gtrsim 200 \, \rm{GeV}$. Hence we conclude that a heavy Higgs boson is disfavored by the EWPT also in the SCSM.

\section{Conclusions}
\label{concl}
In this work we have considered the Strongly Coupled Standard Model, an alternative theory for the electroweak interactions. The low-energy theory, under the assumptions discussed above, is indistinguishable from the conventional Standard Model. Since the $SU(2)_{L}$ is assumed confining rather than spontaneously broken, all the left-handed physical particles are composite states bound into $SU(2)_{L}$ singlets, and a rich spectrum of resonances should eventually emerge. The scale where we expect that to happen is the characteristic scale of the confining interaction, namely the Fermi scale $G_{F}^{-1/2}$, which is within the range of the Large Hadron Collider. While we wait for the direct searches at the LHC we can constrain the model by studying the contributions of the excited states to the electroweak observables and by comparing them with the experimental data.

We have considered one particular resonance of the spectrum, the first radial excitation of the $W$ bosons. Such a triplet is the most constrained state by the EWPT, since its introduction violates the VMD hypothesis, which is expected to be a very good approximation. Indeed we find the effective couplings for this triplet to be very constrained by the experiments, implying that if the model is correct such states have to be rather decoupled from the minimal sector. In particular the VMD hypothesis cannot be violated by more than 1\%. The smallness of the effective parameters makes the model quite unnatural, even though the inclusion of other excited states might alter the analysis, making the allowed region larger and the couplings more natural.

In the Strongly Coupled Standard Model the physics at the Fermi scale is governed by a strong dynamics, which does not allow us to directly compute any quantity which can be compared with the experiments. The best we can do is the effective Lagrangian approach adopted in this work, by writing all the operators consistent with the symmetries of the model and then constraining the effective couplings by a comparison with the experimental data. If the LHC does find new particles with the same quantum numbers of the $W$ and the $Z$ boson the SCSM should be considered as a possible explanation. In addition the model predicts bound states which cannot be viewed as excitations of any Standard Model field, like fermion-fermion bound states. They can be classified in diquark, dileptons and leptoquarks, and they might also be observed at the LHC, giving a very characteristic signature.

\section{Acknowledgments}
I am very thankful to Edward Farhi for suggesting to me this problem, for many useful discussions and for a careful reading of the manuscript. I would also like to thank Andrea De Simone, Massimiliano Procura, Iain Stewart and Jesse Thaler for useful conversations, and Tommaso Borghi, Andreas Hocker, Marco Micheli and Frank Tackmann for their help concerning the numerical computations. This work is supported by U.S. Department of Energy (D.O.E.) under cooperative research agreement DE-FG0205ER41360.

\appendix

\section{Diagonalization of the effective Lagrangian}
\label{appdiag}
In this appendix we show explicitly how to diagonalize the effective Lagrangian. We make the diagonalization for two cases: only minimal sector, minimal sector and the excited triplet.
\subsection{Vector-bosons in the minimal sector}
The effective Lagrangian for the charged vector-bosons is already diagonal, and the correspondent mass eigenstates are $W^{\pm}_{\mu}=\frac{W^1_{\mu}\mp W^2_{\mu}}{\sqrt{2}}$, each one with mass $m_{W}$. \\
The quadratic Lagrangian for the neutral vector-bosons is
\begin{equation}
\mathcal{L}_{quad} =  - \frac{1}{4} W^{3\,\mu\nu} W_{\mu\nu}^3 - \frac{1}{4} \mathcal{F}_{\mu\nu}\mathcal{F}^{\mu\nu} + \frac{1}{2}m_{W}^2 W^{3}_{\mu} W^{3\,\mu} - \frac{k}{2} \mathcal{F}^{\mu\nu} W_{\mu\nu}^3
\end{equation}
It is convenient to rewrite this Lagrangian in the matrix form
\begin{equation}
\mathcal{L}_{quad} =  - \frac{1}{4} \Psi_{\mu\nu}^{T} \cdot K \cdot \Psi_{\mu\nu} + \frac{1}{2} \Psi_{\mu}^{T} \cdot M^2 \cdot \Psi_{\mu}
\end{equation}
where we define the column vectors $\Psi$ as
\begin{equation}
\Psi_{\mu\nu}= \left(\begin{array}{c} \mathcal{F}_{\mu\nu}\\
W^{3}_{\mu\nu}\end{array}
\right) \;\;\;\;\;\;\;\;\;\;\;\;\;\;\; \Psi_{\mu}=\left(\begin{array}{c} a_{\mu}\\
W^{3}_{\mu}\end{array}\right)
\end{equation}
and the matrices
\begin{equation}
K=\left(\begin{array}{cc}
1 & k \\
k & 1\end{array}
\right) \;\;\;\;\;\;\;\;\;\;\;\;\;\; M^2=\left(\begin{array}{cc}
0 & 0 \\
0 & m_{W}^2\end{array}
\right)
\end{equation}

We perform the diagonalization in two steps: we first diagonalize and normalize the matrix $K$ of the kinetic term by performing a non orthogonal transformation, and then we diagonalize the resulting mass term. Let $V_1$ be the matrix which normalizes the kinetic term
\begin{equation}
V_1^{T} \cdot K \cdot V_1 = \left(\begin{array}{cc}
1 & 0 \\
0 & 1\end{array}\right)
\end{equation}
It is possible to show that this matrix is
\begin{equation}
V_1 = \left(\begin{array}{cc}
\frac{1}{\sqrt{1-k^2}} & 0 \\
-\frac{k}{\sqrt{1-k^2}} & 1\end{array}\right)
\end{equation}
If we now redefine the fields as $\Psi = V_1 \cdot \Phi$ the Lagrangian as a function of the new fields becomes
\begin{equation}
\begin{split}
\mathcal{L}_{quad} =&  - \frac{1}{4} \Psi_{\mu\nu}^{T} \left(V_1^{-1}\right)^{T} V_1^{T}\cdot K \cdot V_1 \cdot V_1^{-1} \Psi_{\mu\nu} + \frac{1}{2} \Psi_{\mu}^{T} \cdot M^2 \cdot \Psi_{\mu} = \\
& - \frac{1}{4} \Phi_{\mu\nu}^{T} \cdot \Phi_{\mu\nu} + \frac{1}{2} \Phi_{\mu}^{T} \cdot V_1^{T} \cdot M^2 \cdot V_1  \cdot \Phi_{\mu}
\end{split}
\end{equation}
The kinetic term is now diagonalized and normalized, therefore we have to take care only of the symmetric mass term and diagonalize it by performing an orthogonal transformation.

The mass matrix is now $M_1^2= V_1^{T} \cdot M^2 \cdot V_1$. We have to diagonalize it. Let $V_2$ the orthogonal matrix which performs this transformation
\begin{equation}
V_2^{T} \cdot M_1^2 \cdot V_2 = \left(\begin{array}{cc}
m_{a}^2 & 0 \\
0 & m_{b}^2\end{array}\right)
\end{equation}
The matrix $V_2$ results in
\begin{equation}
V_2 = \left(\begin{array}{cc}
\sqrt{1-k^2} & k \\
k & \sqrt{1-k^2}
\end{array}\right)
\end{equation}
It is important to check that it is an orthogonal matrix: $V_2^{T}\cdot V_2 = 1$. The eigenvalues are
\begin{equation}
M_{d}^2= V_2^{T} \cdot M_1^2 \cdot V_2 = \left(\begin{array}{cc}
0 & 0 \\
0 & \frac{m_{W}^2}{1-k^2}\end{array}\right)
\end{equation}
If we define the new fields $\Phi = V_2 \cdot \Theta$ we can finally rewrite the Lagrangian in the diagonal form
\begin{equation}
\begin{split}
\mathcal{L}_{quad} =&  - \frac{1}{4} \Phi_{\mu\nu}^{T} \cdot \Phi_{\mu\nu} + \frac{1}{2} \Phi_{\mu}^{T} \cdot  \left(V_2^{-1}\right)^{T} V_2^{T}\cdot M_1^2 \cdot V_2 \cdot V_2^{-1}  \cdot \Phi_{\mu} = \\
&  - \frac{1}{4} \Theta_{\mu\nu}^{T} \cdot \Theta_{\mu\nu} + \frac{1}{2} \Theta_{\mu}^{T} \cdot M_{d}^2 \cdot \Theta_{\mu}
\end{split}
\end{equation}
The fields $\Theta_{\mu}$ are the mass eigenstates of the theory. Their expression as a function of the initial fields results in
\begin{equation}
\Theta_{\mu}=\left(\begin{array}{c} A_{\mu}\\
Z_{\mu}
\end{array}\right) =
V_2^{-1} V_1^{-1} \Psi_{\mu}=
\left(\begin{array}{cc}
1 & k \\
0 & \sqrt{1-k^2}
\end{array}\right)
\left(\begin{array}{c} a_{\mu}\\
W^{3}_{\mu}\end{array}\right)
\end{equation}

In conclusion the change of basis is
\begin{equation}
\left(\begin{array}{c} A_{\mu}\\
Z_{\mu}
\end{array}\right) =\left(\begin{array}{cc}
1 & k \\
0 & \sqrt{1-k^2}
\end{array}\right)
\left(\begin{array}{c} a_{\mu}\\
W^{3}_{\mu}\end{array}\right)
\label{changebasisSCSM}
\end{equation}

\subsection{Adding the excited triplet}
The effective Lagrangian for the excited charged vector-bosons is already diagonal, the correspondent mass eigenstates are $W^{\prime \pm}_{\mu}=\frac{W^{\prime 1}_{\mu}\mp W^{\prime 2}_{\mu}}{\sqrt{2}}$, each one with mass $m_{W^{\prime}}$. There is again mixing only in the neutral sector, where we have a quadratic part
\begin{equation}
\begin{split}
\mathcal{L}_{quad} =&  - \frac{1}{4} W^{3\,\mu\nu} W_{\mu\nu}^3 - \frac{1}{4} \mathcal{F}_{\mu\nu}\mathcal{F}^{\mu\nu} + \frac{1}{2}m_{W}^2 W^{3}_{\mu} W^{3\,\mu} - \frac{k}{2} \mathcal{F}^{\mu\nu} W_{\mu\nu}^3\\
& - \frac{1}{4} W^{\prime 3\,\mu\nu} W_{\mu\nu}^{\prime 3} + \frac{1}{2}m_{W^{\prime}}^2 W^{\prime 3}_{\mu} W^{\prime 3\,\mu} - \frac{k^{\prime}}{2} \mathcal{F}^{\mu\nu} W_{\mu\nu}^{\prime 3}
\end{split}
\end{equation}
As we did before for the SCSM case we rewrite the Lagrangian in the matrix form
\begin{equation}
\mathcal{L}_{quad} =  - \frac{1}{4} \Psi_{\mu\nu}^{T} \cdot K \cdot \Psi_{\mu\nu} + \frac{1}{2} \Psi_{\mu}^{T} \cdot M^2 \cdot \Psi_{\mu}
\end{equation}
where now the column vectors $\Psi$ are
\begin{equation}
\Psi_{\mu\nu}= \left(\begin{array}{c} \mathcal{F}_{\mu\nu}\\
W^{3}_{\mu\nu}\\
W^{\prime 3}_{\mu\nu}\end{array}
\right) \;\;\;\;\;\;\;\;\;\;\;\;\;\;\; \Psi_{\mu}=\left(\begin{array}{c} a_{\mu}\\
W^{3}_{\mu}\\
W^{\prime 3}_{\mu}\end{array}\right)
\end{equation}
and the matrices
\begin{equation}
K=\left(\begin{array}{ccc}
1 & k & k^{\prime}\\
k & 1 & 0\\
k^{\prime} & 0 & 1\end{array}
\right) \;\;\;\;\;\;\;\;\;\;\;\;\;\; M^2=\left(\begin{array}{ccc}
0 & 0 & 0 \\
0 & m_{W}^2 & 0\\
0 & 0 & m_{W^{\prime}}^2\end{array}
\right)
\end{equation}

The diagonalization process is again divided in two steps. Let $V_1$ be the matrix which normalizes the kinetic term
\begin{equation}
V_1^{T} \cdot K \cdot V_1 = \left(\begin{array}{ccc}
1 & 0 & 0 \\
0 & 1 & 0 \\
0 & 0 & 1\end{array}\right)
\end{equation}
It is possible to show \cite{Ishida:1991dw} that this matrix is
\begin{equation}
V_1 = \left(\begin{array}{ccc}
\frac{1}{\sqrt{1-k^2-k^{\prime 2}}} & 0 & 0 \\
-\frac{k}{\sqrt{1-k^2-k^{\prime 2}}} & 1 & 0 \\
-\frac{k^{\prime}}{\sqrt{1-k^2-k^{\prime 2}}} & 0 & 1
\end{array}\right)
\end{equation}
If we now redefine the fields as $\Psi = V_1 \cdot \Phi$ the Lagrangian as a function of the new fields becomes
\begin{equation}
\begin{split}
\mathcal{L}_{quad} =&  - \frac{1}{4} \Psi_{\mu\nu}^{T} \left(V_1^{-1}\right)^{T} V_1^{T}\cdot K \cdot V_1 \cdot V_1^{-1} \Psi_{\mu\nu} + \frac{1}{2} \Psi_{\mu}^{T} \cdot M^2 \cdot \Psi_{\mu} = \\
& - \frac{1}{4} \Phi_{\mu\nu}^{T} \cdot \Phi_{\mu\nu} + \frac{1}{2} \Phi_{\mu}^{T} \cdot V_1^{T} \cdot M^2 \cdot V_1  \cdot \Phi_{\mu}
\end{split}
\end{equation}

We have to diagonalize now the mass matrix $M_1^2= V_1^{T} \cdot M^2 \cdot V_1$. Let $V_2$ the orthogonal matrix which performs this transformation
\begin{equation}
V_2^{T} \cdot M_1^2 \cdot V_2 = \left(\begin{array}{ccc}
m_{a}^2 & 0 & 0 \\
0 & m_{b}^2 & 0 \\
0 & 0 & m_{c}^2 \end{array}\right)
\end{equation}
The matrix $V_2$ is a quite complicated expression, but it is anyway possible to obtain an analytical expression for it. As usual a good check is to control that we get an orthogonal matrix $V_2^{T}\cdot V_2 = 1$, and of course to check also that the resulting mass matrix is diagonal. The physical fields now are given by the relation $\Phi = V_2 \cdot \Theta$. We finally rewrite the Lagrangian in the mass eigenstate basis
\begin{equation}
\begin{split}
\mathcal{L}_{quad} =&  - \frac{1}{4} \Phi_{\mu\nu}^{T} \cdot \Phi_{\mu\nu} + \frac{1}{2} \Phi_{\mu}^{T} \cdot  \left(V_2^{-1}\right)^{T} V_2^{T}\cdot M_1^2 \cdot V_2 \cdot V_2^{-1}  \cdot \Phi_{\mu} = \\
&  - \frac{1}{4} \Theta_{\mu\nu}^{T} \cdot \Theta_{\mu\nu} + \frac{1}{2} \Theta_{\mu}^{T} \cdot M_{d}^2 \cdot \Theta_{\mu}
\end{split}
\end{equation}
The fields $\Theta_{\mu}$ are the mass eigenstates of the theory
\begin{equation}
\Theta_{\mu}=\left(\begin{array}{c} A_{\mu}\\
Z_{\mu}\\
Z^{\prime}_{\mu}
\end{array}\right) =
V_2^{-1} V_1^{-1}
\left(\begin{array}{c} a_{\mu}\\
W^{3}_{\mu}\\
W^{\prime 3}_{\mu}
\end{array}\right)
\end{equation}
and the inverse transformation is
\begin{equation}
\left(\begin{array}{c} a_{\mu}\\
W^{3}_{\mu}\\
W^{\prime 3}_{\mu}
\end{array}\right)
= V_1 V_2\left(\begin{array}{c} A_{\mu}\\
Z_{\mu}\\
Z^{\prime}_{\mu}
\end{array}\right) = V \left(\begin{array}{c} A_{\mu}\\
Z_{\mu}\\
Z^{\prime}_{\mu}
\end{array}\right)
\end{equation}
where we define the transformation matrix $V=V_1 V_2$. Once we find the expression for $V$ it is possible to show the following facts which simplify our analysis
\begin{equation}
a_{\mu} = A_{\mu} - k W^{3}_{\mu} -
k^{\prime} W^{\prime 3}_{\mu}; \hspace{3cm} V_{21}=V_{31}=0
\end{equation}
This helps us in rewriting the change of basis in a nicer form
\begin{equation}
\begin{split}
a_{\mu} = & A_{\mu} - k W^{3}_{\mu} -
k^{\prime} W^{\prime 3}_{\mu}\\
W^{3}_{\mu}= & \alpha_{1} Z_{\mu} + \alpha_{2} Z_{\mu}^{\prime}\\
W^{\prime 3}_{\mu}=& \alpha_{3} Z_{\mu} + \alpha_{4} Z_{\mu}^{\prime}
\end{split}
\end{equation}
with the identifications $\alpha_1=V_{22}$, $\alpha_2=V_{23}$, $\alpha_3=V_{32}$, $\alpha_4=V_{33}$. Thus we can limit to consider this subspace. As explained in the text it is reasonable to make an expansion around the zero-th order solution found in equation (\ref{changebasisSCSM}), corresponding to the case $\epsilon=0$ and $\mu=0$ (no excited triplet). We define
\begin{equation}
\alpha_1 = \frac{1}{c_{w}} + \delta_1 \hspace{2cm} \alpha_2 = \delta_2 \hspace{2cm} \alpha_3 = \delta_3 \hspace{2cm} \alpha_4 = 1 + \delta_4
\end{equation}
where the solutions are up to quadratic order in $\epsilon$
\begin{equation}
\begin{split}
\delta_1=&  \frac{\left(-2 s_{w}^4 \mu + 2 s_{w}^6 \mu + s_{w}^4 \mu^2\right)}{2c_{w}^3\left(c_{w}^2-\mu\right)^2} \epsilon^2 \\
\delta_2=& \frac{s_{w}^2}{c_{w}^2-\mu} \epsilon\\
\delta_3=&-\frac{s_{w}^2}{c_{w}\left(c_{w}^2-\mu\right)}\mu \epsilon\\
\delta_4=& \frac{s_{w}^2-s_{w}^4-2 s_{w}^2\mu+ s_{w}^2\mu^2}{2\left(c_{w}^2-\mu\right)^2} \epsilon^2
\end{split}
\end{equation}

\section{Observables used in the global fit}
\label{appobs}
In this appendix we list the observables used in the global fit, and we express them as a function of their Standard Model predictions and the parameters $\left(\epsilon,\mu,\lambda\right)$.

\subsection*{W mass}
The predicted $W$ mass is
\begin{equation}
m_{W}^2 = m_{Z}^2 \left[c_{w}^2+\frac{s_{w}^4}{c_{w}^2-\mu}\mu\epsilon^2\right]
\end{equation}

\subsection*{Partial and total Z decay width}
The partial decay width for the process $Z \rightarrow f \bar{f}$ can be computed by using the Lagrangian in the equation (\ref{LNCZ}) and it results in
\begin{equation}
\Gamma_{f} = N_{c}^{f}\, m_{Z}\,\frac{\alpha}{6 s_{w}^2 c_{w}^2} \left(\left|g_{L}^{f}\right|^2+\left|g_{R}^{f}\right|^2\right)
\end{equation}
where $N_{C}$ is the color factor for the fermion $f$. The overall normalization is now changed by the new states, and from equation (\ref{sw2}) we have
\begin{equation}
\frac{1}{s_{w}^2 c_{w}^2}= \frac{1}{s_{0}^2 c_{0}^2} \frac{\left(1+\epsilon\lambda\right)^2}{\left(1 + \mu \lambda^2\right)} \left(1+\frac{s_{w}^4}{c_{w}^2\left(c_{w}^2-\mu\right)}\mu\epsilon^2\right)
\end{equation}
We can separate the contribution due to the excited states by expanding about the SM value
\begin{equation}
\Gamma_{f} = \Gamma_{f}^{SM} \left[1 + 2 \epsilon\lambda - \mu \lambda^2 + \frac{s_{w}^4}{c_{w}^2\left(c_{w}^2-\mu\right)}\mu\epsilon^2 + 2 \frac{g_{L}^{f, SM} \delta g_{L}^{f} + g_{R}^{f, SM} \delta g_{R}^{f}}{\left(g_{L}^{f, SM}\right)^2+\left(g_{R}^{f, SM}\right)^2}\right]
\end{equation}

The observables we are going to consider for our global fit are the total Z width
\begin{equation}
\Gamma_{Z} = \sum_{f,\,f\neq t} \Gamma_{f}
\end{equation}
where the sum runs over all the SM fermions but not the top quark, and the branching ratios
\begin{equation}
R_{e} = \frac{\Gamma_{had}}{\Gamma_{e}}; \hspace{1cm} R_{\mu} = \frac{\Gamma_{had}}{\Gamma_{\mu}}; \hspace{1cm} R_{\tau} = \frac{\Gamma_{had}}{\Gamma_{\tau}}; \hspace{1cm} R_{b} = \frac{\Gamma_{b}}{\Gamma_{had}}; \hspace{1cm} R_{c} = \frac{\Gamma_{c}}{\Gamma_{had}}
\end{equation}
where $\Gamma_{had}$ is the partial Z width in hadrons. In addition we consider also the hadronic cross section at the $Z$ pole
\begin{equation}
\sigma^{h}_{p} = 12\pi \frac{\Gamma_{e} \Gamma_{had}}{m_{Z}^2 \Gamma_{Z}^2}
\end{equation}

\subsubsection*{Left-Right asymmetry}
The expression for the left-right asymmetry is
\begin{equation}
A_{LR} = \left[\frac{g_{L}^{e\,2}-g_{R}^{e\,2}}{g_{L}^{e\,2}+g_{R}^{e\,2}}\right]
\end{equation}
If we assume the deviation from the SM couplings to be small we can rewrite this relation as
\begin{equation}
A_{LR} = A_{LR}^{SM} + 4 \frac{g_{L}^{e, SM}g_{R}^{e, SM}}{\left[\left(g_{L}^{e, SM}\right)^2+\left(g_{R}^{e, SM}\right)^2\right]^2}\left[g_{R}^{e, SM}\delta g^{e}_{L}- g_{L}^{e, SM}\delta g^{e}_{R}\right]
\end{equation}

\subsubsection*{Forward-backward asymmetries}
We want to evaluate the unpolarized forward-backward asymmetry for the scattering $e^{+} e^{-} \rightarrow \bar{f} f$. We have to distinguish between two cases: when the fermions in the final states are a lepton pair $l^{+} l^{-}$ (where $l= e, \mu, \tau$) we have the relation
\begin{equation}
A_{FB}^{l^{+} l^{-}} = \frac{3}{4} A_{LR}^{e^{+} e^{-}} A_{LR}^{l^{+} l^{-}}
\end{equation}
whereas when they are a heavy quark pair $\bar{q} q$ (where $q= c, b$) we have the relation
\begin{equation}
A_{FB}^{\bar{q} q} = \frac{3}{4} \left(1-k_{A}\frac{\alpha_{s}}{\pi}\right)A_{LR}^{e^{+} e^{-}} A_{LR}^{\bar{q} q}
\end{equation}
The factor $\left(1-k_{A}\frac{\alpha_{s}}{\pi}\right)$ is a radiative QCD correction and can be taken to be 0.93 \cite{Burgess:1993vc}. In our fit we include: $A_{FB}^{e^{+} e^{-}}$, $A_{FB}^{\mu^{+} \mu^{-}}$, $A_{FB}^{\tau^{+} \tau^{-}}$, $A_{FB}^{\bar{b} b}$, and $A_{FB}^{\bar{c} c}$.

\subsubsection*{$\tau$ asymmetries}
The polarization asymmetry $A_{pol}(\tau)$ for the process $e^{+} e^{-} \rightarrow \tau \bar{\tau}$ is defined by
\begin{equation}
A_{pol}(\tau) = \frac{\sigma_{R} - \sigma_{L}}{\sigma_{R} + \sigma_{L}} = - \left[\frac{g_{L}^{\tau\,2}-g_{R}^{\tau\,2}}{g_{L}^{\tau\,2}+g_{R}^{\tau\,2}}\right] \equiv - \mathcal{A}_{\tau}
\end{equation}
where $R,L$ is the helicity of the final state. In our fit we use  $\mathcal{A}_{\tau}$, which results in
\begin{equation}
\mathcal{A}_{\tau} = \mathcal{A}_{\tau}^{SM} + 4 \frac{g_{L}^{\tau, SM}g_{R}^{\tau, SM}}{\left[\left(g_{L}^{\tau, SM}\right)^2+\left(g_{R}^{\tau, SM}\right)^2\right]^2}\left[g_{R}^{\tau, SM}\delta g^{\tau}_{L}- g_{L}^{\tau, SM}\delta g^{\tau}_{R}\right]
\end{equation}
It is possible to measure also the joint forward-backward/left-right asymmetry for the same process, which is defined as
\begin{equation}
A_{e}\left(P_{\tau}\right) = \frac {\sigma^{\tau}_{LF}-\sigma^{\tau}_{LB}-\sigma^{\tau}_{RF}+\sigma^{\tau}_{RB}}{\sigma^{\tau}_{LF}+\sigma^{\tau}_{LB}+\sigma^{\tau}_{RF}+\sigma^{\tau}_{RB}} = \frac{3}{4} \left[\frac{g_{L}^{e\,2}-g_{R}^{e\,2}}{g_{L}^{e\,2}+g_{R}^{e\,2}}\right] \equiv \frac{3}{4} \mathcal{A}_{e(\tau)}
\end{equation}
and again by expanding $\mathcal{A}_{e(\tau)}$ about the SM value we find
\begin{equation}
\mathcal{A}_{e(\tau)} = \mathcal{A}_{e(\tau)}^{SM}  + 4 \frac{g_{L}^{e, SM}g_{R}^{e, SM}}{\left[\left(g_{L}^{e, SM}\right)^2+\left(g_{R}^{e, SM}\right)^2\right]^2}\left[g_{R}^{e, SM}\delta g^{e}_{L}- g_{L}^{e, SM}\delta g^{e}_{R}\right]
\end{equation}

\section{Experimental data and Standard Model predictions}
\label{appGFITTER}
In this appendix we list the observables used to make the electroweak fit, with their experimental values and the SM predictions. The experimental data are taken from \cite{:2005ema,Alcaraz:2006mx,Collaboration:2008ub}, whereas the Standard Model predictions have been computed by GFITTER \footnote{We thank Andreas Hocker for providing us GFITTER and for his help.} \cite{Flacher:2008zq}. We compute the observables for $m_h=115\,\rm{GeV}$ and $m_{t}=172.4 \pm 1.2 \,\rm{GeV}$ \cite{:2008vn}.
\begin{table}[H]
\begin{center}
\begin{tabular}{|c|c|c|}\hline
Quantity & Experiment & SM($m_h=115\,\rm{GeV}$) \\ \hline
$m_W \, \rm{(GeV)}$ & 80.399 $\pm$ 0.025 & 80.360 \\
$\Gamma_Z \, \rm{(GeV)}$ & 2.4952 $\pm$ 0.0023 & 2.4944 \\
$R_e$ & 20.804 $\pm$ 0.050 & 20.731 \\
$R_\mu$ & 20.785 $\pm$ 0.033 & 20.731 \\
$R_\tau$ & 20.764 $\pm$ 0.045 & 20.731 \\
$R_b$ & 0.21629 $\pm$ 0.00066 & 0.2158  \\
$R_c$ & 0.1721 $\pm$ 0.0030 & 0.1722 \\
$\sigma_h \, \rm{(nb)}$ & 41.541 $\pm$ 0.037 & 41.486 \\
$A_{LR}$ & 0.1513 $\pm$ 0.0021 & 0.1469 \\
$A_{FB}^{e^+ e^-}$ & 0.0145 $\pm$ 0.0025 & 0.0162 \\
$A_{FB}^{\mu^+ \mu^-}$ & 0.0169 $\pm$ 0.0013 & 0.0162 \\
$A_{FB}^{\tau^+ \tau^-}$ & 0.0188 $\pm$ 0.0017 & 0.0162 \\
$A_{FB}^{\bar{b} b}$ & 0.0992 $\pm$ 0.0016 & 0.1030 \\
$A_{FB}^{\bar{c} c}$ & 0.0707 $\pm$ 0.0035 & 0.0736 \\
$\mathcal{A}_{\tau}$ & 0.1439 $\pm$ 0.0043 & 0.1469 \\
$\mathcal{A}_{e(\tau)}$ & 0.1498 $\pm$ 0.0049 & 0.1469 \\
\hline
\end{tabular}
\end{center}
\caption{Observables used to make the EW fit}
\label{tableOBS}
\end{table}

\bibliographystyle{apsrev}
\bibliography{Wprime}

\end{document}